\documentclass{article}
\usepackage{spconf,amsmath,graphicx}
\usepackage{float}

\title{DuDe : Dual-Decoder Multilingual ASR for Indian Languages using Common Label Set}
\name{A Arunkumar$^{*}$, Mudit Batra$^{*}$, S Umesh\thanks{*These authors have contributed equally.}}
\address{Speech Lab, Indian Institute of Technology Madras}
\begin{document}
%
\maketitle
\begin{abstract}
In a multilingual country like India, multilingual Automatic Speech Recognition (ASR) systems have much scope. Multilingual ASR systems exhibit many advantages like scalability, maintainability, and improved performance over the monolingual ASR systems. However, building multilingual systems for Indian languages is challenging since different languages use different scripts for writing. On the other hand, Indian languages share a lot of common sounds. Common Label Set (CLS) exploits this idea and maps graphemes of various languages with similar sounds to common labels. Since Indian languages are mostly phonetic, building a parser to convert from native script to CLS is easy. In this paper, we explore various approaches to build multilingual ASR models. We also propose a novel architecture called Encoder-Decoder-Decoder for building multilingual systems that use both CLS and native script labels. We also analyzed the effectiveness of CLS-based multilingual systems combined with machine transliteration.
\end{abstract}
\section{Introduction}
\label{sec:intro}

India is a country with great linguistic diversity having 22 constitutionally recognized languages. So, it becomes important to build Automatic Speech Recognition (ASR) systems for the regional languages in India. Most of the Indian languages are considered to be low resource since the amount of labelled speech data available is less. This makes it challenging to build robust ASR systems for Indian languages. In recent times, there has been increased interest in building multilingual ASR systems\cite{Kannan2019, 9053443, kumar21e_interspeech, sailor21_interspeech, joshi21_interspeech, 9414961} for several reasons. Firstly, a multilingual model is easy to maintain since it replaces multiple monolingual models with a single model. Secondly, the multilingual models benefit from more data obtained by combining various languages. Traditionally, in machine learning and deep learning, more data helps in better parameter estimation leading to better performance. Multilingual models attempt to exploit the same idea and achieve better performance through parameter sharing across languages.

Unlike other language families like Latin languages, different Indian languages use different scripts for writing. This makes it more challenging to build multilingual ASR systems for Indian languages. More specifically, this hinders the models from learning the common sounds across the languages. Despite different writing scripts, Indian languages share a rich amount of common sounds. Motivated by this, a common label set for Indian languages was proposed for speech synthesis\cite{ramani13_ssw}. A rule-based unified parser for the conversion of native language text to phoneme-based CLS was proposed in \cite{10.1007/978-3-319-45510-5_59}. CLS was shown to work well for TTS in \cite{ramani13_ssw}. In \cite{kumar21e_interspeech} and \cite{9053443}, CLS was shown to give some improvement for ASR. \cite{9053443} and \cite{Thomas2020} used a very similar idea to CLS in which they used transliterated text across languages to train multilingual ASR systems.

This work explores various approaches including language-dependent and language independent multilingual models. This work mainly focuses on using the Common Label Set (CLS) for building multilingual ASR systems. This work also proposes a novel architecture for building a language independent multilingual ASR system that might pave way for future research in this area.

The rest of the paper is organized as follows. Section \ref{sec:data} briefs the dataset used in this work. Section \ref{sec:CLS} describes the Common Label Set. Section \ref{sec:proposed} details various approaches we followed to build a multilingual ASR system, including the proposed method. Section \ref{sec:results} discusses the results of various experiments performed in this work. Section {sec:conclusion} draws important conclusions based on the results observed.

\section{Dataset Details}
\label{sec:data}
The dataset used in this paper is Shrutilipi\cite{shruti} which is a labeled ASR corpus obtained by mining parallel audio and text pairs at the document scale from All India Radio news bulletins for 12 Indian languages. The Indo-Aryan languages, namely Hindi, Gujarati, Marathi, Bengali, and Odia, are considered in this paper. The audio files present in the dataset are sampled at $16$KHz frequency. A random subset of 150 hours is selected for each language as the training data. Further, a validation set and an evaluation set, each of 2 hours, are sampled from the given corpus.

\section{Common Label Set}
\label{sec:CLS}
\begin{figure}[H]
    \centering
    \includegraphics[scale=0.65]{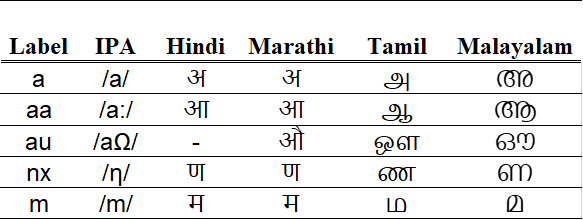}
    \caption{Examples of CLS Mapping with Native Script Characters}
    \label{fig:cls-1}
\end{figure}

\begin{figure}[H]
    \centering
    \includegraphics[scale=0.7]{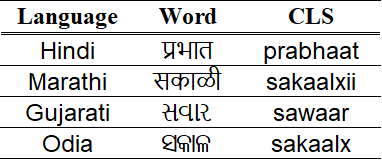}
    \caption{Examples of CLS equivalent of Native Script words}
    \label{fig:cls-2}
\end{figure}

The acoustic similarity among the same set of phones of different languages suggests the possibility of a compact and common set of labels. The common label set assigns a standard set of labels for speech sounds that occur in Indian languages. Since Indian languages have a high correlation between graphemes and phonemes, grapheme to phoneme conversion can be easy and rule-based. This work uses the CLS proposed in \cite{ramani13_ssw}. The similar sounds in different languages are mapped to the same label. Each CLS label is a sequence of alphanumeric characters of the Roman alphabet. For example, the sound represented by the International Phonetic Alphabet (IPA) /u:/ is represented by the CLS label "uu". Some examples of CLS mapping with some Indian language scripts are given in Fig.\ref{fig:cls-1}. For generating CLS scripts from the corresponding Native Script, Unified Parser\cite{10.1007/978-3-319-45510-5_59} is used. The parser first identifies the language through Unicode range and then uses the language specific rules to output the CLS-based representation. From Fig.\ref{fig:cls-2}, we can see that even though the scripts are different, the similar sounding phones are mapped to the same CLS labels. One important thing to note is that due to certain special cases like schwa deletion, geminate correction, and syllable parsing, reconstructing native script text from CLS text is not straightforward.

\vspace{-1.5em}

\begin{figure*}[t]
\centering
\includegraphics[scale=0.72]{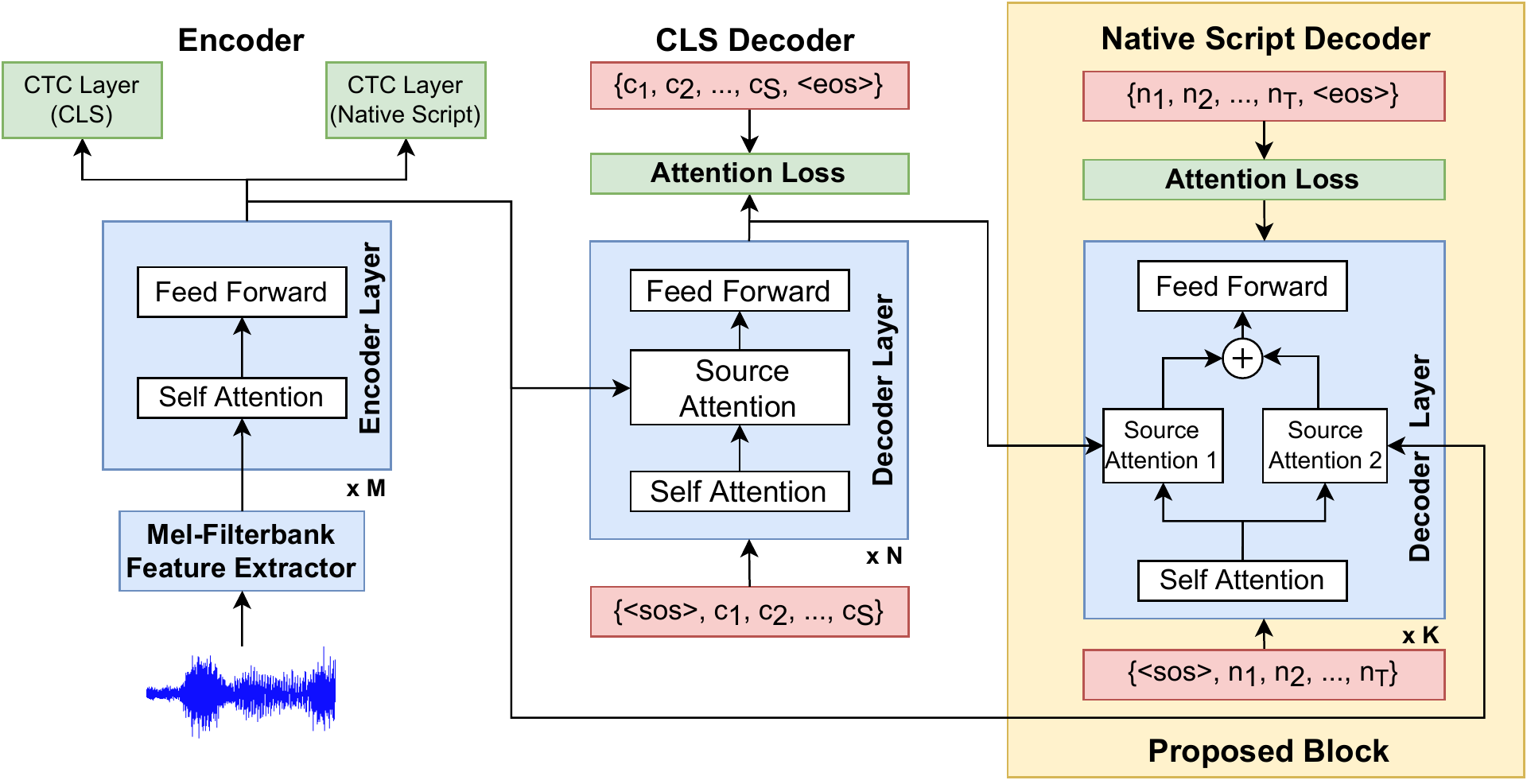}
  \caption{Proposed Encoder-Decoder-Decoder Architecture}
  \label{EDD model}
\end{figure*}

\vspace{-1.5em}
\section{Multilingual ASR models}
All models are trained using the ESPNet\cite{watanabe2018espnet} toolkit. 80 dimensionsal Mel-Filterbank features with 3 pitch features are used as an input for each experiment. The frame size is kept as 25ms with the frame shift of 10ms. The models are trained using Joint CTC-attention \cite{hybrid-ctc} based Transformer architecture\cite{vaswani2017attention}. The common parameters for all models used for the Transformer Network are shown in the Table \ref{tab:General Parameters}.

\begin{table}[h]
\begin{tabular}{c c}
\hline 
Hyperparameters          & Values  \\ \hline \hline
Width          & 512  \\ \hline
Feed Forward Units                     & 2048 \\ \hline
Attention heads                   & 8    \\ \hline
CTC weight                        & 0.3  \\ \hline
\end{tabular}
\centering
\caption{Model Configurations}
\label{tab:General Parameters}
\end{table}

\label{sec:proposed}
\subsection{Approach 1: Vanilla Multilingual E2E model}
In this approach, a vanilla multilingual model is built by pooling data from all the languages during training to produce a multilingual system. The resultant model is language-independent since no language information is passed during training and it serves as a baseline.

A 750-hour training set is obtained by combining the 150-hour training set of all five languages. The combined training set is used to train the model. A transformer model with 15 Encoder layers and 9 Decoder layers is used. The number of BPE units used in this model is 2000.

\subsection{Approach 2: CLS-based Multilingual E2E model}
In order to take advantage of the common sounds across Indian languages, the native script text is converted to CLS text, as explained in section \ref{sec:CLS}. In this approach, an E2E ASR model is trained to predict CLS text, given speech. However, we have to produce outputs in the native script eventually. Once we have a multilingual CLS model, we can do many things. We can either finetune the CLS model with the native script labels or feed the CLS output to a machine transliteration model that will convert CLS to the native script. Both approaches are investigated in this work and explained below.

A transformer model with 15 Encoder layers, 9 Decoder layers, and 1000 BPE units is used to train the multilingual CLS model. Since there are many common sounds across languages, we have reduced the number of BPE units compared to the vanilla multilingual model.

\subsubsection{Multilingual CLS Finetuned for Aryan Language Family}
The multilingual CLS model is finetuned with the combined training set using native script labels. We typically expect this model to be better than the vanilla multilingual one. Since the finetuning is with the native script labels, the number of BPE units is increased to 2000. This model will be language-agnostic.

\subsubsection{Multilingual CLS Finetuned for Individual Languages}
The multilingual CLS model is finetuned separately for each language. This process results in language-specific monolingual models. The number of BPE units is reduced to 450 since the language-specific data is only used for finetuning.

\subsubsection{Multilingual CLS Output Fed Through Machine Transliteration System}
This approach involves building a machine transliteration (MT) model that converts CLS text to native script text. We need excellent transliteration models to exploit the gains realized by the CLS system. Otherwise, the transliteration system might introduce errors that will affect the system's overall performance. A transformer model with 6 encoder layers, 6 decoder layers, and 4 attention heads is used to train the machine transliteration model for each language. The machine transliteration models are trained using Fairseq toolkit\cite{ott2019fairseq}. All machine transliteration models are trained only using the text corresponding to the 150 hours of training data of each language.






\subsection{Approach 3: Encoder-Decoder-Decoder based E2E CLS-Native script model}
This paper proposes a novel, language-independent multilingual ASR model that can give output in both CLS and native script. The block diagram is shown in Fig.\ref{EDD model}. Decoder 1 is learning to predict CLS output, while Decoder 2 is learning to predict the native script output. As we already mentioned, the CLS combines various Indian languages into a single phonetic space so that the decoder can handle all languages simultaneously. The acoustic information captured by the encoder is passed to both decoders for cross-attention. Additionally, the native script decoder attends to the output of the CLS decoder. With the proposed model, we are attempting to integrate the machine transliteration model into a single E2E model by adding a dual attention decoder while getting the benefits of a CLS-based model.

This proposed method can directly have output from the Native script Decoder without language identification. Another approach in this setup is to tap the CLS output from CLS Decoder and pass it through the machine transliteration model. 

The proposed model is trained with 12 encoder layers, 6 CLS decoder layers, and 6 native script decoder layers. The total number of layers in both systems is the same, i.e., 24, to allow for an fair comparison with the Multilingual E2E baseline. For ease of implementation, the number of BPE units is 2000 for both CLS and native script.

\begin{table*}[t]
\centering
\footnotesize
\setlength\tabcolsep{4pt}
\begin{tabular}{c c c c c c c c}
\hline
Model & Architecture detail & Hindi & Gujarati & Marathi & Bengali & Odia & Average\\ \hline \hline
\begin{tabular}[c]{@{}c@{}}Monolingual\\ (Baseline)\end{tabular}       & \begin{tabular}[c]{@{}c@{}}12+6 (450 BPE)\end{tabular}           & 28.9  & 40.5     & 23.0    & 36.9    & 28.7 & 31.6\\ \hline
\begin{tabular}[c]{@{}c@{}}Multilingual\\ (Baseline)\end{tabular}      & \begin{tabular}[c]{@{}c@{}}15+9 (2000 BPE)\end{tabular}          & 29.7  & 40.2     & 22.6    & 37.2    & 27.9 & 31.5\\ \hline
\begin{tabular}[c]{@{}c@{}}Multilingual CLS\\ + MT model\end{tabular}  & \begin{tabular}[c]{@{}c@{}}15+9 (1000 BPE)\end{tabular}          & \textbf{28.3}  & \textbf{36.5}     & \textbf{17.6}    & \textbf{31.3}    & \textbf{22.4} & \textbf{27.2} \\ \hline
\begin{tabular}[c]{@{}c@{}} Proposed \\ Multilingual NS \end{tabular}  & \begin{tabular}[c]{@{}c@{}}12+6+6 (2000 BPE)\end{tabular} & 30.3 & 40.8     & 23.5    & 38.2    & 28.5 & 32.3\\ \hline
\begin{tabular}[c]{@{}c@{}} Proposed Multilingual CLS\\ + MT model\end{tabular}  & \begin{tabular}[c]{@{}c@{}}12+6+6 (2000 BPE)\end{tabular}   & 30.0 & 37.0     & 18.4    & 32.4    & 23.1 & 28.2\\ \hline
\end{tabular}
\caption{WER (\%) comparison of various Multilingual ASR models.}
\label{tab:Baseline Comparison}
\end{table*}

\begin{table*}[h]
\centering
\footnotesize
\setlength\tabcolsep{4pt}
\begin{tabular}{c c c c c c c c}
\hline
Model & Architecture detail & Hindi & Gujarati & Marathi & Bengali & Odia & Average\\ \hline \hline

\begin{tabular}[c]{@{}c@{}}Multilingual CLS\\ Fine-tuned (all)\end{tabular}      & \begin{tabular}[c]{@{}c@{}}15+9 (2000 BPE)\end{tabular}          & 29.1  & 40.2     & 22    & 36.2    & 27.6 & 31.0\\ \hline
\begin{tabular}[c]{@{}c@{}}Multilingual CLS\\ Fine-tuned (solo)\end{tabular}   & \begin{tabular}[c]{@{}c@{}}15+9 (450 BPE)\end{tabular}          & 29.1  & 39.8     & 22    & 36.3    & 27.9 & 31.0\\ \hline
\begin{tabular}[c]{@{}c@{}} Proposed Multilingual CLS\\ Fine-tuned (solo)\end{tabular}  & \begin{tabular}[c]{@{}c@{}}15+9 (450 BPE)\end{tabular}  & 29.6  & 40.6     & 22.8    & 37.0    & 28.3 & 31.6\\ \hline
\begin{tabular}[c]{@{}c@{}}Multilingual CLS\\ + MT model\end{tabular}  & \begin{tabular}[c]{@{}c@{}}15+9 (2000 BPE)\end{tabular}          & \textbf{28.3}  & \textbf{36.5}     & \textbf{17.6}    & \textbf{31.3}    & \textbf{22.4} & \textbf{27.2} \\ \hline
\end{tabular}
\caption{WER(\%) comparison of various Finetuned Multilingual ASR models with CLS-MT mode.}
\label{tab:Baseline finetune Comparison}
\end{table*}

\vspace{-0.5em}
\section{Results and Discussion}
\label{sec:results}
We begin by comparing the outcomes of the Multilingual baseline model and the Monolingual baseline model. Table.\ref{tab:Baseline Comparison} demonstrates that although data for multilingual systems is five times greater than that of monolingual systems, there is no discernible benefit. As Indo-Aryan languages use different scripts, it becomes difficult for the model to learn all languages with one model. To overcome this issue, we built a CLS-based model whose output is passed to the Machine Transliteration model. Compared to the Monolingual Baseline, there has been a 13.9\% relative WER improvement. The CLS model takes advantage of the fact that all languages are placed in the same phonetic space, and switching from CLS to native script using the MT model becomes a simple process.

We also explored the Encoder-Decoder-Decoder-based technique, as shown in Fig.\ref{EDD model}, to create a universal multilingual model that could output both CLS and native scripts using a single model. The benefit of this model is that we need only a single Native script decoder to handle it, eliminating the need to create different MT models for each language. However, in Table\ref{tab:Baseline Comparison}, we can see a relative decline in WER of 2.47\% over the Multilingual baseline. The reasoning behind the declining performance is that despite receiving information from the encoder and CLS decoder, the native script decoder still experiences confusion while producing output in the native script.

In another approach with Encoder-Decoder-Decoder Model, we tap CLS output and pass it to the MT model. We expected that the CLS output from this model might be more effective than the Multilingual CLS baseline model since the Native Script decoder guides it during training. We observe that this approach gave a relative improvement of 10.75\% WER over the Monolingual Baseline but a drop of 1\% as compared to the Multilingual CLS + MT model.

Another way to go from the CLS domain to the native script is to finetune the models with the native script labels. Once again, two strategies are tested. Firstly, finetune the multilingual CLS model with each language separately. Secondly, finetune the multilingual CLS model with the combined training set to create an Indo-Aryan Multilingual model. From the Table.\ref{tab:Baseline finetune Comparison}, it can be observed that finetuning in any way is not helping the CLS model to create a better native script model. We also tried finetuning experiments on the proposed Encoder-Decoder-Decoder architecture. However, for a fair comparison with other finetuning experiments, we had to train a proposed model with 15 encoder layers, 9 CLS decoder layers, and 6 Native Script decoder layers. Encoder and only CLS Decoder are taken, again finetuned with Native scripts individually. From this result, it is again observed that finetuning is not a good idea compared to passing CLS output to the MT model. Also, training machine transliteration models is easy and less resource intensive compared to finetuning.

\vspace{-1em}
\section{Conclusion}
\label{sec:conclusion}
This paper analyzed various approaches for developing a Multilingual ASR system that uses the Common Label Set. The CLS maps the common sounds across the Indian languages to the same label. The experiments show that passing the CLS output to the Machine transliteration system is far superior to other methods. The proposed novel Encoder-Decoder-Decoder architecture for Multilingual ASR gives output in both the CLS domain and Native script and is language-independent. Though its performance is not competitive, the model is language-independent and has a lot of scope for further research. The CLS multilingual model cascaded with the machine transliteration model gave us the best relative improvement of 13.9\% WER over the Monolingual baseline.

\section{Acknowledgement}
\label{sec:ack}
A part of this work was funded by "Bhashini: National Language translation Mission" project of the Ministry of Electronics and Information Technology  (MeitY), Government of India.


\newpage
\bibliographystyle{IEEEbib}
\bibliography{strings,refs}

\end{document}